\documentclass[aps,prl,amsmath,amssymb, reprint, superscriptaddress,showpacs]{revtex4-1}
\usepackage{color, graphicx}     % Include figure files
\usepackage{dcolumn}     % Align table columns on decimal point
\usepackage{bm}                  % bold math
\usepackage{amssymb}
\usepackage{latexsym}
\usepackage{amsfonts}
\usepackage{amsmath}
\usepackage{multirow}
\usepackage[hidelinks]{hyperref}  % Para incluir hiperkinks
\usepackage{hyperref}
\usepackage[dvipsnames]{xcolor}%\usepackage{changes}
\usepackage{times}
\begin{document}
%\linenumbers
%\nolinenumbers
%\preprint{APS/123-QED}

%\usepackage{graphicx}

\makeatletter
\newcommand*\bigcdot{\mathpalette\bigcdot@{.3}}
\newcommand*\bigcdot@[2]{\mathbin{\vcenter{\hbox{\scalebox{#2}{$\m@th#1\bullet$}}}}}
\makeatother

\title
{{Anomalous enhancement of thermal radiation transport by quasidisorder}}

\newcommand*{\HIT}[0]{{School of Energy Science and Engineering, Harbin Institute of Technology, Harbin 150001, China}}
\newcommand*{\KLAT}[0]{{Key Laboratory of Aerospace Thermophysics, Ministry of Industry and Information Technology, Harbin, China}}
\newcommand*{\IUF}[0]{{Institut Universitaire de France, 1 rue Descartes, F-75231 Paris, France}}
\newcommand*{\LCC}[0]{{Laboratoire Charles Coulomb (L2C), UMR 5221 CNRS-Université de Montpellier, F-34095 Montpellier, France}}
\newcommand*{\NUS}[0]{{Department of Electrical and Computer Engineering, National University of Singapore, Singapore 117583, Singapore}}

\author{Cheng-Long Zhou}
\affiliation{\HIT}%
\affiliation{\KLAT}%
\author{Xin-Yu Jia}
\affiliation{\HIT}%
\affiliation{\KLAT}%
\author{Shui-Hua Yang}
\affiliation{\NUS}%
\author{Yan Wang}
\affiliation{\HIT}%
\affiliation{\KLAT}%
\author{Yong Zhang}
\affiliation{\HIT}
\affiliation{\KLAT}%
\author{Hong-Liang Yi}
\email{Corresponding author: yihongliang@hit.edu.cn}
\affiliation{\HIT}%
\affiliation{\KLAT}%
\author{Mauro Antezza}
\email{Corresponding author: mauro.antezza@umontpellier.fr}
\affiliation{\IUF}
\affiliation{\LCC}%

%\date{\today}        
\begin{abstract}
%	a$\bigcdot$b
The transition from order to disorder is conventionally regarded as detrimental to solid-state heat transfer in classical wave and quasiparticle systems. In striking contrast, we show that in near-field thermal radiation, breaking long-range order—shifting from periodic to quasiperiodic configurations—induces a counterintuitive enhancement of energy transport. This effect arises from delocalized interactions within quasiperiodic elements, where this quasiperiodicity relays and amplifies thermal electromagnetic energy transfer across large spatial separations, surpassing even corresponding near-field scenarios. The extraordinary transport properties induced by quasidisorder effect of near-field thermal radiation could unlock exciting opportunities for heat flow manipulation, offering transformative implications for thermal science and advancing the fundamental understanding of collective excitations in non-ordered systems.

\end{abstract}
\maketitle
%------------------------------------------------------------------------------------------------Sec I
\setlength{\parskip}{0pt}
\textit{Introduction.}--The transition from order to disorder is a fundamental paradigm in condensed matter physics, profoundly shaping the optical\,\textcolor{black!40!blue}{\cite{Engin1995order,John1987LOCALIZATION,White2009Bosons}}, electronic\,\textcolor{black!40!blue}{\cite{Baryam1986ELECTRONIC,Kane2005spin,Fu2007insulators}}, mechanical\,\textcolor{black!40!blue}{\cite{Salmani2020,Agarwal2020Curved,Lerner2022Glassy}}, and thermal properties \,\textcolor{black!40!blue}{\cite{LiBW2001,Roychowdhury2021}} of materials. This structural breaking underlies a wealth of extraordinary phenomena, including phase transitions\,\textcolor{black!40!blue}{\cite{Wall2018}}, charge-density waves\,\textcolor{black!40!blue}{\cite{Bordia2016}}, and emergent topological states\,\textcolor{black!40!blue}{\cite{ZhangWX2021}}. Notably, chaos and quasidisorder give rise to unique quantum effects, including enhanced soliton transport\,\textcolor{black!40!blue}{\cite{Martinez2008}}, Bose glass formation\,\textcolor{black!40!blue}{\cite{Ciardi2023}}, and Mott insulating states\,\textcolor{black!40!blue}{\cite{YaoHP2024}}. Extensive mapping of phase diagrams for interacting bosons has been validated experimentally in one- and two-dimensional quasidisordered systems\,\textcolor{black!40!blue}{\cite{Ristivojevic2012,Bertoli2018}}, while the discovery of superconductivity in quasicrystals also underscores the pivotal role of the order-to-disorder transition in revealing critical phenomena and collective excitations\,\textcolor{black!40!blue}{\cite{Uri2023,LiuYB2024}}. Beyond its foundational role, this transition provides a versatile framework for designing materials with tailored functionalities, enabling breakthroughs in fields spanning acoustic metamaterials to optoelectronic device\,\textcolor{black!40!blue}{\cite{YangZJ2015,Tennyson2019}}. 

\begin{figure*}[t]
	\centering
	\centerline{\includegraphics[width=1\textwidth]{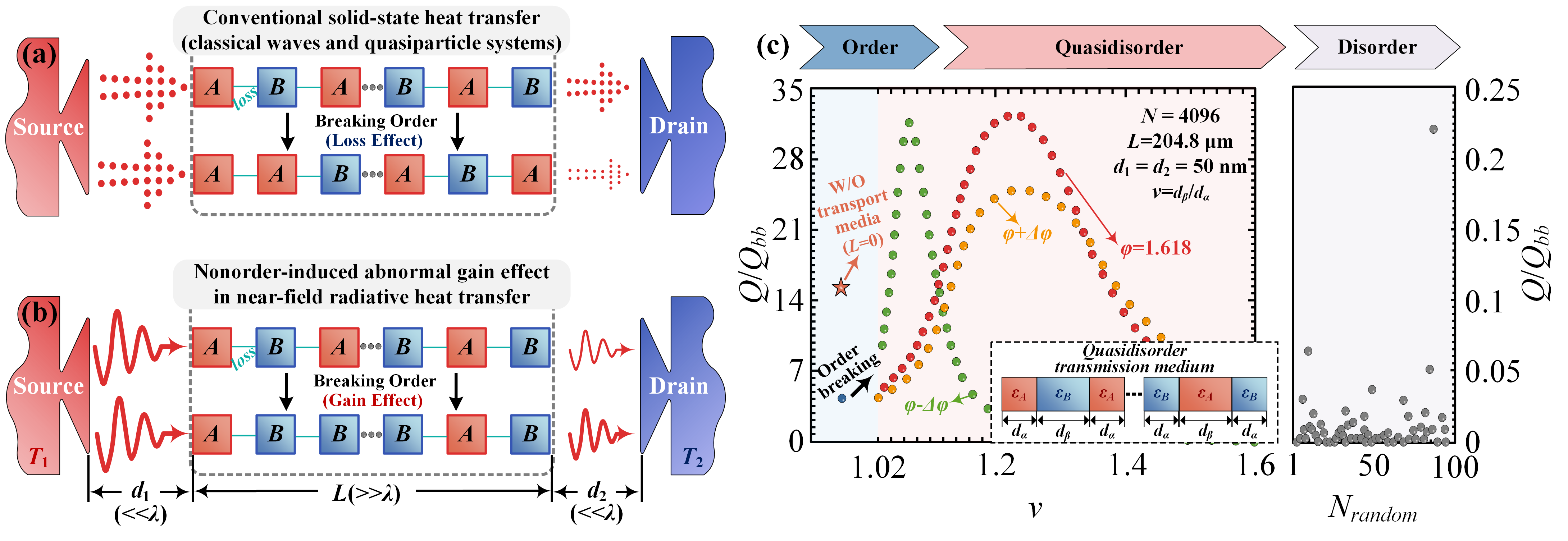}}
	\caption{(a) Schematics of the conventional solid-state heat transfer before (after) the order-breaking. (b) Schematics of  near-field thermophotonic transport enhancement induced by the transition from order to non-ordered state. The thermal photons of source are transmitted to the drain via transport medium with length \textit{L}. Between both reservoirs and the intermediate slab is a vacuum gap of thickness $d$. (c) The evolutionary trajectory of radiative heat flux $Q$ during transition of the transport medium from an ordered state to a quasidisordered state and subsequently to a chaotic state. The five-pointed star indicate the heat flux between source and drain without transport medium ($L=0$). $Q_{bb}=\sigma (T_{1}^4-T_{2}^4)$ represents Stephen Boltzmann's law, where $\sigma$ is Stephen Boltzmann constant. Structural details are available in the Supplemental Material \,\textcolor{black!40!blue}{\cite{SM}}.}
	\label{Fig1}
\end{figure*}

Similarly, breaking the atomic and subwavelength structural order can significantly impact heat transfer mechanisms\,\textcolor{black!40!blue}{\cite{Pohl2002,Abanin2019}}. However, unlike convective systems that can optimize thermal efficiency by generating turbulence through non-ordered dynamics\,\textcolor{black!40!blue}{\cite{Howard1963,Busse1969}}, breaking spatial order in classical wave and quasiparticle systems typically suppresses solid-state heat transport\,\textcolor{black!40!blue}{\cite{Savic2008,ChangCW2008,Garg2011,Vardeny2013,Coppolaro2018,Coppolaro2020}} [Fig.\textcolor{black!40!blue}{~\ref{Fig1}(a)}]. In crystalline materials, non-ordered state introduces additional phonon scattering leading to phonon localization and the consequent suppression of heat flow\,\textcolor{black!40!blue}{\cite{Savic2008}}. Likewise, in thermophotonic systems, disorder and quasidisorder affect electromagnetic coherence and coupling, resulting in Anderson-type localization effects and spectral discretization that severely hinder the  energy transport of propagative wave \,\textcolor{black!40!blue}{\cite{Coppolaro2020,Wiersma1997,Lahini2008,YuS2021}}. This raises a fundamental question: could there exist solid-state heat transfer mechanisms where breaking spatial order instead facilitates energy transport? Probing such order-breaking-driven heat transfer behaviors in solid media could deepen our understanding of collective excitations in complex systems and provide transformative insights into thermal technologies. Yet, harnessing disorder or quasidisorder to enhance solid-state heat transfer remains an unresolved challenge.

In this letter, we present a near-field thermophotonic system with spatial quasidisorder, revealing, contrary to conventional solid-state heat transfer paradigms, an anomalous behavior of near-field thermophotonic energy transfer during the transition from order to disorder [Fig.\textcolor{black!40!blue}{~\ref{Fig1}(b)}]. Using fluctuational electrodynamics (FED)\,\textcolor{black!40!blue}{\cite{zhang2007nano,carminati1999near,Shen2009nl}} and the Green's function approach (GFA)\,\textcolor{black!40!blue}{\cite{biehs2021near,Francoeur2009,volokitin2007near}}, we uncover that breaking spatial order within the thermophotonic transport pathway induces delocalized interactions with significant gain effects. Furthermore, we identify heat flow oscillation emerging from order breaking and establish its phenomenological scaling laws exhibiting strong nonlinear characteristics. These discoveries challenge the prevailing notion that order-breaking in solid-state heat transfer invariably leads to losses\,\textcolor{black!40!blue}{\cite{HuR2020}}, instead demonstrating a novel mechanism where order-breaking robustly enhances heat transfer in solid medium. Additionally, we uncover a nontrivial spatial dependence in quasidisorder-induced transport enhancement, where evanescent energy transport across far-field distances can surpasses its near-field baseline by severalfold, deviating significantly from the conventional exponential decay law of near-field thermophotonics. These findings suggest that thermophotonic quasidisorder transport mechanism could serve as a promising platform for energy transport, exhibiting unprecedented efficiencies surpassing the efficiency of many traditional thermophotonic transfer modes proposed to date.  

As illustrated in Fig.\textcolor{black!40!blue}{~\ref{Fig1}(b)}, the input energy is initially excited by the source, positioned on one side of the transport medium. Our goal is to analyse the heat flux through the transmission system, where the source is maintained at a temperature of $T_{1}=301$ K, while the drain is at $T_{2}=300$ K. To ensure the accuracy of our concept demonstration in this work, the temperature of the transfer medium is strictly kept equal to that of the drain. The thermal photon transport calculations follow the standard FED procedure\,\textcolor{black!40!blue}{\cite{Rodriguez2011PRL}} and the GFA method\,\textcolor{black!40!blue}{\cite{Miller2015PRL}}. The radiative heat flux (RHF) within the drain is expressed as\,\textcolor{black!40!blue}{\cite{Qu2024}}:
\begin{align}
	\label{eq:1}
	\begin{aligned}
		Q=&\int_{0}^{\infty} q(\omega) d\omega=\int_{0}^{\infty}d\omega\int_{-\infty}^{\infty}\frac{d\textbf{k}}{(2\pi)^{3}}4\omega\epsilon_{0}\\
		&\times (\Theta(\omega, T_{1})-\Theta(\omega, T_{2}))\\
		&\times Re\left\{Tr\left[\hat{\Gamma}\hat{G}_{E}(k,\omega)\frac{\hat{\epsilon}-\hat{\epsilon}^{\dagger}}{2i} \hat{G}^{\dagger}_{H}(k,\omega)\right]\right\}
	\end{aligned}
\end{align}
where $q(\omega)$ is the spectral heat transfer coefficient. $\Theta(\omega, T) = \hbar\omega/(e^{\hbar\omega/k_{b}T} - 1)$ represents the average energy of a photon at frequency $\omega$ and temperature $T$, and $k_{b}$ is the Boltzmann constant. $\hat{G}_{E}(\hat{G}_{H})$ denotes the electric (magnetic) Green's function, $k$ is the in-plane wavevector, and $\epsilon$ is the permittivity tensor of the source. In this study, both the source and drain are composed of silicon carbide. The matrix $\hat{\Gamma}$ is defined as $\left[0\;-1\;0;1\;0\;0;0\;0\;0 \right]$. Further computational details are available in the Supplemental Material\,\textcolor{black!40!blue}{\cite{SM}}. We consider a generic transport medium system comprising an array of two propagating modes ($A$ and $B$). The array consists of $N$ objects. In an ordered system, the two modes, $A$ and $B$, alternate with equal propagation distances, $d_{0}$. A quasidisordered structure emerges between order and complete disorder when this alternating sequence is absent. In this work, we mainly examine the impact of Fibonacci geometry, a classical quasidisordered system, on thermophotonic transfer. Fibonacci geometry is generated through iterative inflation rules\,\textcolor{black!40!blue}{\cite{Albuquerque2004}}: $\alpha \to \alpha\beta, \beta \to \alpha$, associating the distances $d_{\alpha}$ and $d_{\beta}$ with symbols $\alpha$ and $\beta$, respectively. Alternatively, a cut-and-project approach can directly calculate the mode distribution\,\textcolor{black!40!blue}{\cite{Albuquerque2004}}:
\begin{equation}
	\label{eq:2}
     z_{n}=d_{\alpha}{\parallel{\frac{n}{\varphi}}\parallel}+d_{\beta}(n-\parallel{\frac{n}{\varphi}}\parallel),
\end{equation}
here, $\varphi$ is the ratio between the number of $\alpha$ and $\beta$ symbols  ($\varphi=N_{\alpha}/N_{\beta}$). And, it can be shown that, in the asymptotic limit of an infinite sequence, the ratio between the number of symbols $\alpha$ and $\beta$ approaches the Golden Mean\,\textcolor{black!40!blue}{\cite{Albuquerque2004}}. In Eq.\textcolor{black!40!blue}{~\ref{eq:3}}, we used

\begin{figure}[t]
	\centering
	\centerline{\includegraphics[width=1\columnwidth]{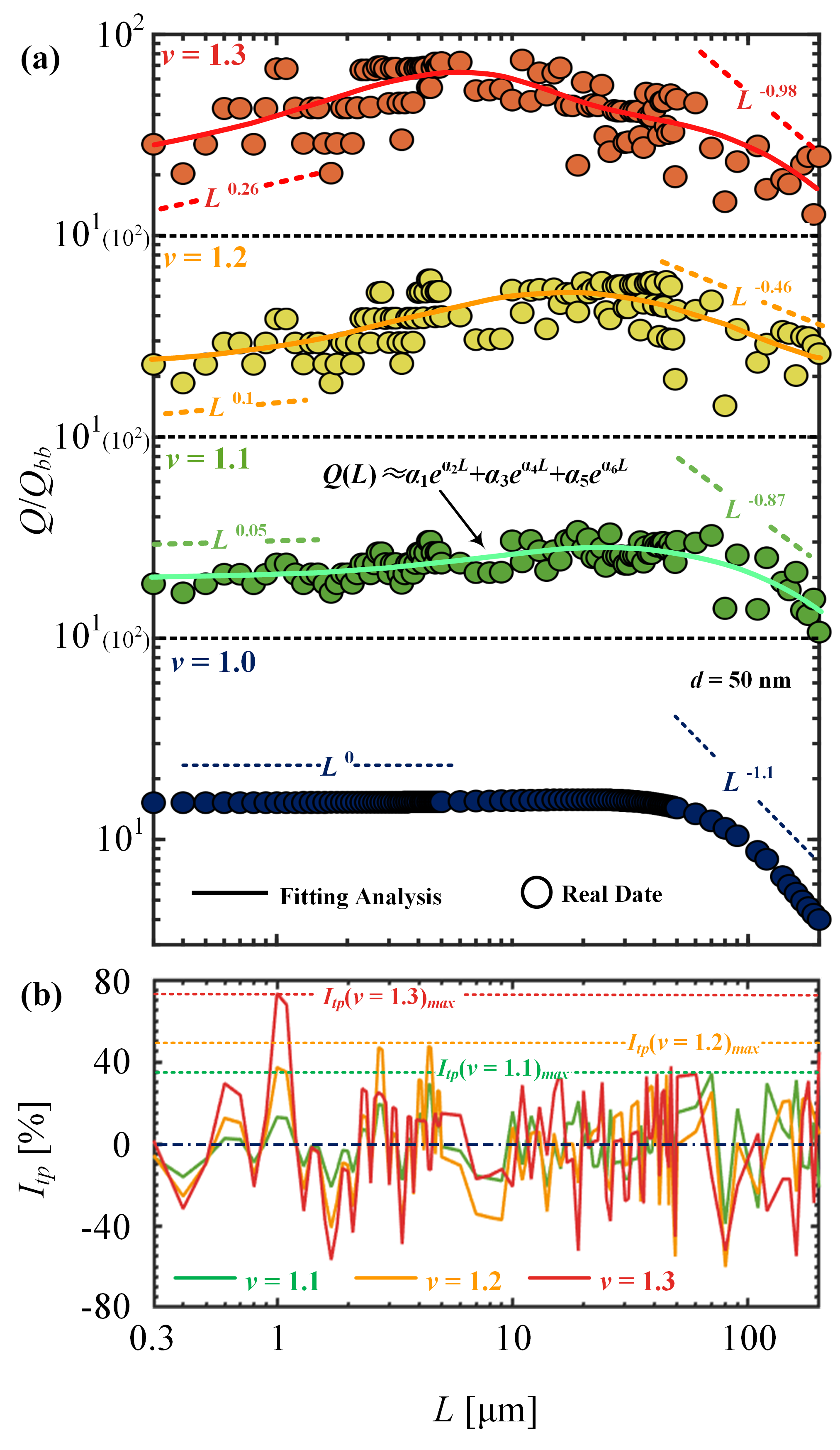}}
	\caption{(a) Heat transfer from source to drain as a function of $L$ for different quasidisorder degree, and compared to cases of the order at the same parameters. Points show  numerically exact results [computed using Eq.\textcolor{black!40!blue}{~\ref{eq:1}}], while solid lines represent the approximate trend (see SM\,\textcolor{black!40!blue}{\cite{SM}} for parameter). (b) Heat fluctuating ratio of this radiative system with increaning $L$ for different quasidisorder degree.}
	\label{Fig2}
\end{figure}

\begin{equation}
	\label{eq:3}
\parallel{x}\parallel=\left\{
\begin{array}{l}
	n,\quad\;\;\;  n\;{\leq}\;x\;{\le} \;n+\frac{1}{2}\\
	n+1,n\;+\;\frac{1}{2}\;{\leq}\;x\;{\leq}\;n+1
\end{array}
\right.,
\end{equation}
where, $n$ is the $n$-th body an array of Fibonacci geometry. It is important to note that, unlike typical Fibonacci-type geometries in the literature\,\textcolor{black!40!blue}{\cite{Albuquerque2004}}, our approach assumes only that propagation distances are distributed according to the Fibonacci sequence, while propagation modes alternate. This design implies that each cell in the system has four possible combinations of propagation mode and distance. This modified scheme facilitates comparisons with effective medium theory predictions as well as with periodic reference structures. To quantify the degree of quasidisorder, we define the scale-ratio parameter:
\begin{equation}
	\label{eq:4}
    v=\frac{d_{\beta}}{d_{\alpha}}, \quad\;\;\;1\;\leq\;v\;\le\infty.
\end{equation}

By varying the quasidisorder index $v$, we analyze the transition from perfect periodicity ($v = 1$) to different levels of quasiperiodic order ($v \geq 1$). Within this framework, the average layer thickness is defined as $d_{0} = L/N$, where $L$ represents the total propagation distance (see Fig.\textcolor{black!40!blue}{~\ref{Fig1}(b)}). Using the result in Eq.\textcolor{black!40!blue}{~\ref{eq:4}}, it can be shown that, in the asymptotic limit of an infinite sequence:
\begin{equation}
	\label{eq:5}
	d_{\alpha}=\frac{1+\varphi}{v+\varphi}d_{0}.
\end{equation}

Moreover, to avoid introducing extraneous interferences, this study assumes negligible loss of energy transfer between these two modes of propagation (i.e., adjacent objects are in close adherence), and the array is oriented perpendicular to the infinite propagation direction. Details of the model can be found in the supplementary material\,\textcolor{black!40!blue}{\cite{SM}}. The propagating modes $A$ and $B$ are designated as a vacuum and a perfect resonance body (dielectric constant is -1), respectively. In periodic array systems, solid-state heat transfer has shown remarkable advancements, particularly concerning phonon coherence and superdiffusive behavior\,\textcolor{black!40!blue}{\cite{ChenG2021}}. Similarly, in near-field thermophotonics, periodic arrays exhibit exceptional capabilities for manipulating thermal radiation. Figure \textcolor{black!40!blue}{~\ref{Fig1}(c)} illustrates the pronounced effect of periodic array configurations on radiative heat transfer. The vacuum gap $d_{1}(d_{2})$ is fixed at 50~nm. For propagation distances exceeding a few hundred micrometers ($L=204.8~\mu$m), the radiative heat flux (RHF) approaches nearly four times the blackbody limit.

\textit{Heat transfer singularity of quasidisorder radiative system.}--When the order of a periodic array is perturbed and the system becomes quasidisordered, the heat flux is significantly enhanced, as shown in Fig.\textcolor{black!40!blue}{~\ref{Fig1}(c)}. Intriguingly, this anomalous amplification contrasts starkly with the reduction in phonon thermal conductivity typically induced by order-breaking. In the golden ratio Fibonacci geometry, the RHF can reach a maximum enhancement of up to 32 times the blackbody radiation intensity. This enhancement is robust, even under perturbations of $\Delta\varphi=\pm10\%\varphi$, as indicated in Fig.\textcolor{black!40!blue}{~\ref{Fig1}(c)}. Analogous to Howard's renowned hypothesis\,\textcolor{black!40!blue}{\cite{Howard1963}} $\--$ which posits that turbulence characterized by order-breaking optimizes convective heat transfer $\--$ our findings suggest that quasidisorder may represent an optimal configuration for thermophotonic transfer systems. Notably, a phenomenon of thermal current pulsation has been observed in the quasi-disordered transport of thermal photons. As illustrated in Fig.\textcolor{black!40!blue}{~\ref{Fig2}(a)}, it can be seen that the RHF behavior of the quasidisordered system exhibits pronounced discretization with increasing propagation distance. The propagation distance is extended by increasing $N$. This discretization becomes more evident as the \textit{v} increases. To better analyse the observed discretization behavior, we derive an approximate scaling law for the radiative heat flux of this thermophotonic quasidisorder transport mechanism, 
\begin{equation}
	\label{eq:5}
	Q_{asl}(L)=a_{1}e^{a_{2}L}+a_{3}e^{a_{4}L}+a_{5}e^{a_{6}L},
\end{equation}
elucidating the approximate scale dependence of the RHF behavior in the quasidisordered system (parameters in SM\,\textcolor{black!40!blue}{\cite{SM}}). Interestingly, at far-field distances ($L$=200 $\mu$m), the introduction of breaking-order instead slows down the decaying trend of near-field thermophotonic energy. To visualize the oscillation feature, Fig.\textcolor{black!40!blue}{~\ref{Fig2}(b)} presents the thermophotonic pulsating level ($I_{tp}=(Q(L)-Q_{asl}(L))/Q_{asl}(L)$), with reference to the pulsating level (the ratio of the mean-square sum of the fluctuating velocity to the time-averaged velocity to represent the magnitude of the pulsatione\,\textcolor{black!40!blue}{\cite{DINGZJ2025}}). It has been observed that the thermophotonic pulsating level increases in proportion to the degree of quasidisorder index.

\begin{figure*}[t]
	\centering
	\centerline{\includegraphics[width=1\textwidth]{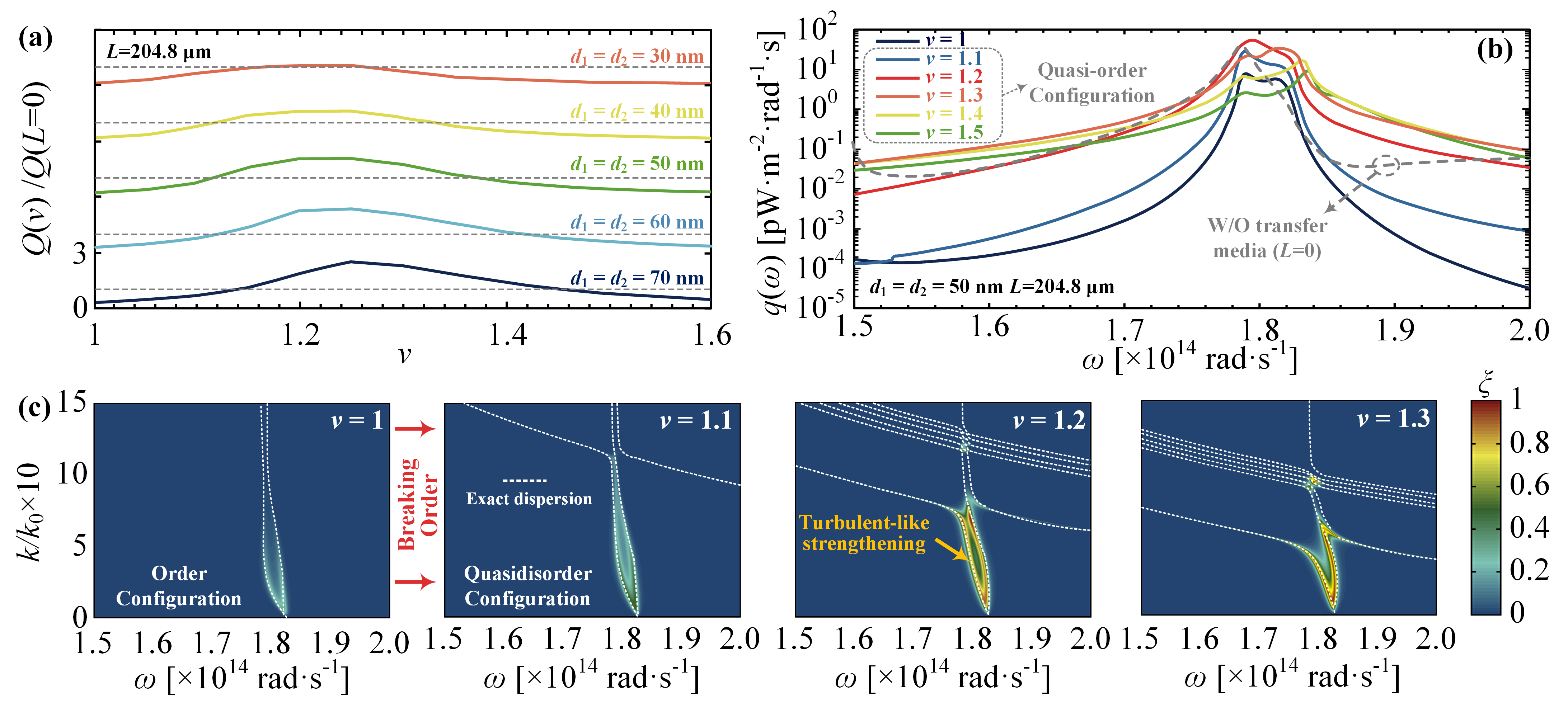}}
	\caption{(a) The ratio of the far-field RHF within quasidisordered system and its corresponding near-field case as a function of quasidisorder index $v$, for different vacuum gaps. The gray dotted line represents $Q(v)/Q(L=0)=1$. (b) The spectral intensity function $q$($\omega$) at different quasidisorder index for $d=50$\,nm. (c) The evolution of thermophotonic probability $\xi$ during the transition from order ($v=1$) to quasidisorder ($v=1.1-1.3$). The white line is dispersion relation of thermophotonic system. The nonlocal analysis can be seen in SM\,\textcolor{black!40!blue}{\cite{SM}}.}
	\label{Fig3}
\end{figure*}

On the other hand, it is well established that far-field heat flux (over hundreds of micrometers) is consistently lower than near-field heat flux (within a few hundred nanometers)\,\textcolor{black!40!blue}{\cite{Tang2024n,Song2015nn,Zhu2019N,rincon2022enhancement,Zhou2024PRL}}. This conclusion holds true for periodic transmission systems, as shown in Fig.\textcolor{black!40!blue}{~\ref{Fig1}(c)}. However, our findings reveal a striking deviation: the RHF for $v \approx 1.2$ can reach up to twice the intensity observed in near-field conditions (without transport media) [see Fig.\textcolor{black!40!blue}{~\ref{Fig1}(c)}]. This counterintuitive phenomenon is remarkably robust: irrespective of the near-field thermophotonic intensity (corresponding to varying gap $d_{1}(d_{2})$), the far-field RHF generated by quasidisordered excitation surpasses its near-field counterpart [see Fig.\textcolor{black!40!blue}{~\ref{Fig3}(a)}].  These findings suggest the presence of complex collective excitations behaviors in strongly disordered thermophotonic systems.

\textit{Thermophotonic quasidisorder transport mechanism.}--The spectral intensity function, which represents the energy distribution of thermophotons across different frequencies, shows a substantial enhancement when transitioning from ordered to quasiordered states, as depicted in Fig.\textcolor{black!40!blue}{~\ref{Fig3}(b)}. As the scale-ratio parameter $v$ increases from 1 to 1.2, the peak spectral intensity rises from $8$ to $56~\mathrm{pW \cdot m^{-2}\cdot{rad}^{-1}\cdot{s}\cdot{K}^{-1}}$, representing a sevenfold increase compared to the corresponding near-field scenario. This significant rise in intensity underpins the anomalous strengthening of thermophotonic energy transfer, as shown in Fig.\textcolor{black!40!blue}{~\ref{Fig1}(c)}. 

To further elucidate this quasidisorder-induced transport enhancement in thermophotonics, we examine the thermophotonic transmission probability $\xi$ (details in SM\,\textcolor{black!40!blue}{\cite{SM}}).\,\,The thermophotonic transmission probability (TTP) quantifies the tunneling probability of thermal photons between emitter and receiver. The dependence of TTP on order-breaking is illustrated in Fig.\textcolor{black!40!blue}{~\ref{Fig3}(c)}. In the ordered case ($v=1$), the TTP retains its narrow-band, large-wavevector characteristics, driven by the phonon-polariton modes of the source, with dispersion closely reflecting these thermophotonic traits. Remarkably, the introduction of the breaking order leads to a significant abundance of the interaction between mode elements,  resulting in increased dispersion modes in radiative transfer [more white dotted lines in Fig.\textcolor{black!40!blue}{~\ref{Fig3}(c)}]. 

It is also not negligible that the order-breaking phenomenon also shows a remarkable enhancement of thermophotonic transmission due to this long-range nonlocal interference effect between quasiperiodic elements.\,\,As can be seen in Fig.\textcolor{black!40!blue}{~\ref{Fig3}(c)}, the TPP intensity can only reach below 0.3 when the transfer medium is ordered. However, when \textit{v} is increased to 1.2, these delocalized interactions enable near-field thermophotons, even after propagating over several hundred micrometers, to generate near-perfect tunneling behavior ($\xi\approx1$) across both broadband and broad-wavevector regimes. Nonlocal analysis methods are provided in the SM\,\textcolor{black!40!blue}{\cite{SM}}.

\begin{figure}[t]
	\centering
	\centerline{\includegraphics[width=1\columnwidth]{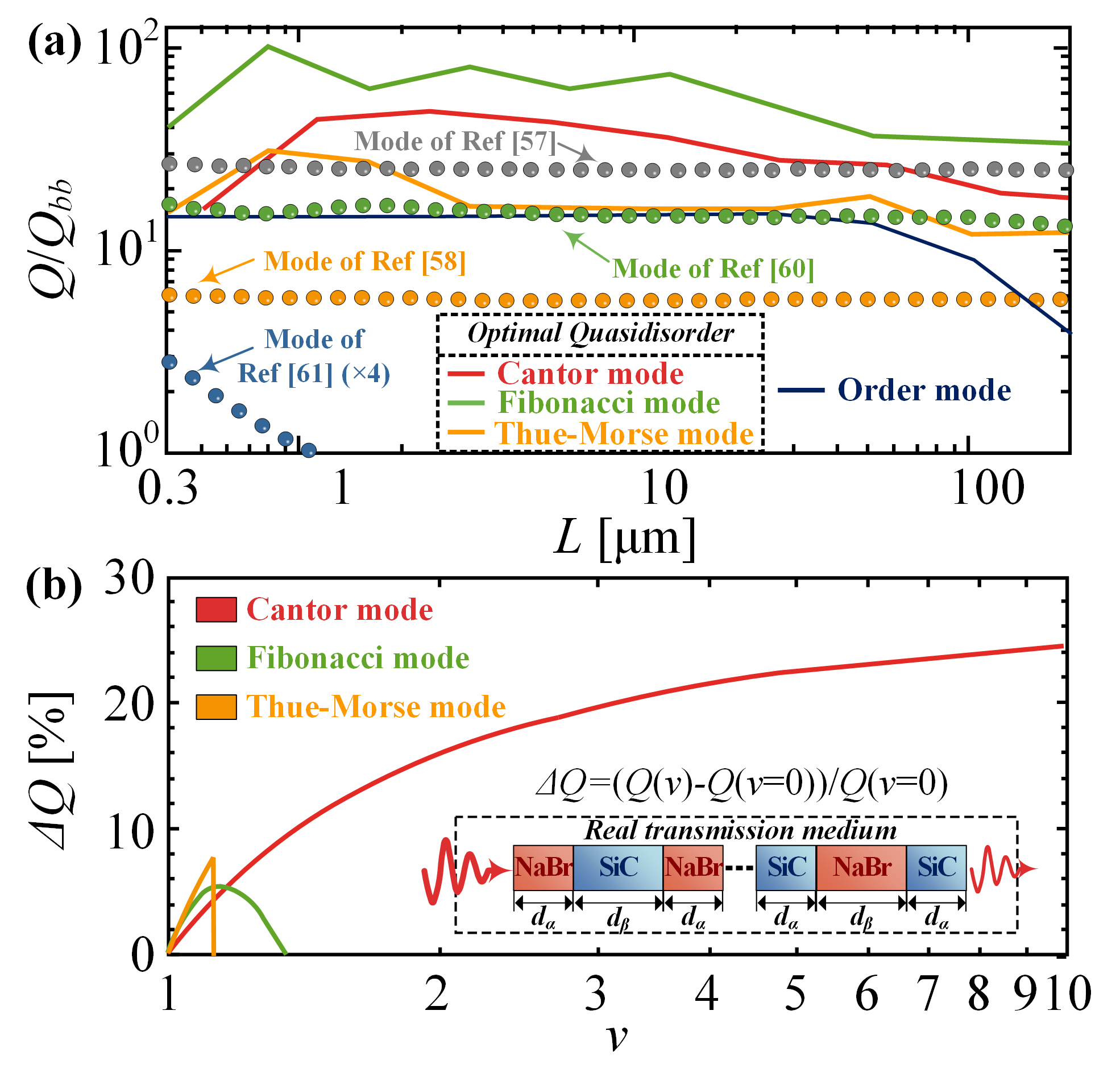}}
	\caption{(a) A comparison of the RHF of quasidisorder transport modes and different modes reported in literatures\,\textcolor{black!40!blue}{\cite{Messina2016,ZhaoB2021,Mirmoosa2017,Asheichyk2022,KanYH2019}}. The RHF of quasidisorder transport modes are corresponding optimal $v$ for the different transmission distance. (b) RHF amplification ratios for different quasidisordered geometries real cell. This real cell consists of NaBr/SiC.}
	\label{Fig4}
\end{figure}

\textit{Super-Planck character of quasidisorder RHF in far-field.}--A key question in thermal photonics research is whether near-field effects can propagate to the far field, thereby enhancing RHF efficiency at large separations\,\textcolor{black!40!blue}{\cite{biehs2021near}}. Numerous solutions have been proposed to address this challenge, including hyperbolic photonic crystals\,\textcolor{black!40!blue}{\cite{Messina2016}}, high-dielectric modes\,\textcolor{black!40!blue}{\cite{ZhaoB2021,Mirmoosa2017}}, nanowire waveguides\,\textcolor{black!40!blue}{\cite{Asheichyk2022}}, and grating waveguides\,\textcolor{black!40!blue}{\cite{KanYH2019}}. However, these mechanisms often fall short in facilitating efficient long-distance near-field energy transmission. In contrast, the quasidisorder transport behavior we report represents a superior mode of energy transfer within the same range of parameters. In Fig.\textcolor{black!40!blue}{~\ref{Fig4}(a)}, we give the thermal transport properties of three classical quasidisordered structure (Fibonacci mode, Thue-Morse mode, and Cantor mode). At a propagation distance of 100 $\mu$m, the RHF is approximately 5 times greater than that achieved by a hyperbolic mechanism and 4 times that of a cylindrical waveguide. This phenomenon is also observed in real quasidisordered systems, such as those formed by NaBr/SiC [Fig.\textcolor{black!40!blue}{~\ref{Fig4}(b)}], offering enhanced thermal manipulation and increased design flexibility for thermal devices.

In summary, we have shown that breaking long-range spatial order—by transforming energy transport pathways from periodic to quasiperiodic configurations—can induce a pronounced enhancement of near-field radiative heat transfer. This nontrivial effect challenges the prevailing expectation that disorder conventionally suppresses thermal transport in solid-state systems. Our analysis reveals that this enhancement originates from strongly delocalized interactions among elements within quasidisordered systems, which amplify collective electromagnetic modes and enable radiative heat transfer surpassing those of periodic structures by up to an order of magnitude. Moreover, this quasidisorder-driven heat flux exhibits remarkable stochastic fluctuations during propagation, reflecting the nonlinearity of delocalized interactions among elements within quasidisordered systems. This thermophotonic quasidisorder transport mechanism not only deepens our fundamental understanding of heat transfer in solid-state media but also provides fresh insights into the dynamics of collective excitations in many-body systems. Beyond its conceptual significance, we have also demonstrated the robustness of the proposed thermophotonically quasidisordered transmission mechanism across diverse quasiperiodic structures and real material combination. This quasidisorder-driven approach offers a highly efficient strategy for transmitting near-field energy across extended distances, surpassing conventional transfer modes and opening avenues for transformative advances in thermal physics, energy conversion, and nanophotonic applications.

\section{Funding}
National Natural Science Foundation of China (Grant Nos.~U22A20210, 523B2060, and 52506075); Postdoctoral Fellowship Program of CPSF (Grant No.~GZB20240951); Funding for Postdoctoral Candidates in Heilongjiang Province (Grant No.~AUGA41100088248).
\section{ACKNOWLEDGMENTS}
This paper acknowledges the discussion from Prof. Zijing Ding of HIT and PhD. Mingqian Yuan of NJU. This work acknowledges support from the National Natural Science Foundation of China (Grant Nos.~U22A20210, 523B2060, and 52506075), the Postdoctoral Fellowship Program of CPSF (Grant No.~GZB20240951), Funding for Postdoctoral Candidates in Heilongjiang Province (Grant No.~AUGA41100088248), the grant ”CAT” from the ANR/RGC Joint Research Scheme sponsored by the French National Research Agency (ANR) (Grant No~A-HKUST604/20), and the Research Grants Council (RGC) of the Hong Kong Special Administrative Region. Part of the computational resources were provided by the DIPC computing center.
\section{Disclosures}
The authors declare no conflict of interest.

\section{Data availability}
The authors declare that all the data and code supporting the findings of this study are available within the article, or upon request from the corresponding author.

\section{Supplemental document}

See Supplement Materials for supporting content.

\bibliography{reference1}

\end{document}